\documentclass[conference,compsoc]{IEEEtran}

\ifCLASSOPTIONcompsoc
  \usepackage[nocompress]{cite}
\else
  \usepackage{cite}
\fi

\ifCLASSINFOpdf
  \usepackage[pdftex]{graphicx}
\else
  \usepackage[dvips]{graphicx}
\fi
\graphicspath{{figures/}}

\usepackage{array}
\usepackage{amsmath}
\usepackage{booktabs}
\usepackage[table]{xcolor}
\usepackage{float}
\usepackage{placeins}
\usepackage{needspace}
\usepackage{stfloats}
\usepackage{flushend}
\usepackage{url}
\usepackage{xspace}
\usepackage[hidelinks]{hyperref}

\newcommand{\paperTitle}{Agent Approval Laundering:\\
Transitive Effects Beyond the Approved Invocation\xspace}
\newcommand{\approval}{approval laundering\xspace}
\newcommand{\finding}[3]{%
  \par\smallskip\noindent
  \begingroup
  \setlength{\fboxsep}{4pt}%
  \fcolorbox{black!25}{black!4}{%
    \parbox{\dimexpr\columnwidth-2\fboxsep-2\fboxrule\relax}{%
      \textbf{Finding #1: #2} #3}}%
  \endgroup
  \par\smallskip}

\title{\paperTitle}

\author{\IEEEauthorblockN{Jinqian Zhang\textsuperscript{1,2},
Haojun Xia\textsuperscript{1,2,$\dagger$},
Shujiang Wu\textsuperscript{3},
Jingkun Yue\textsuperscript{4},\\
Xia Zhang\textsuperscript{1,2},
Zhangpei Cheng\textsuperscript{1,2},
and Bibo Tu\textsuperscript{1,2,$\dagger$}}
\IEEEauthorblockA{\textsuperscript{1}Institute of Information Engineering,
Chinese Academy of Sciences\\
\textsuperscript{2}School of Cyber Security, University of Chinese Academy of
Sciences\\
\textsuperscript{3}Beihang University\\
\textsuperscript{4}State Key Laboratory of Networking and Switching Technology,\\
Beijing University of Posts and Telecommunications, Beijing, China\\
\textsuperscript{$\dagger$}Corresponding authors.\\
Email: zhangjinqian26@mails.ucas.ac.cn, xiahaojun@iie.ac.cn,\\
wushujiang@buaa.edu.cn, yuejk@bupt.edu.cn,\\
zhangxia2025@iie.ac.cn, chengzhangpei26@mails.ucas.ac.cn,\\
tubibo@iie.ac.cn}}

\begin{document}

\maketitle

\begin{abstract}
An approval mechanism can store a durable record that names an approved command
exactly while omitting an effect of the workflow that command starts. We call
this coverage gap \emph{\approval}. The gap is an authorization failure when a
pre-specified effect policy requires different decisions and the complete
policy-visible inputs are identical: no record-only policy can then guarantee
both outcomes. With \texttt{npm run build} held fixed, a change only in the
pre-approval script leaves command semantics allowing both the local script and
the script that fetches. The same transitive workflow can arise when a person runs
the command directly. Coding agents are the measured setting.

We present the first systematic security analysis of this gap as
\emph{closure-bound approval} over six operational effect classes. The classes measure coverage. The Approval-to-Action
Security Benchmark binds captured approval objects and inspectable decision-time
metadata to immutable post-execution evidence.

Across 111 fixed approval-object/trace pairs, residual
records fall from 40 under explicit fields to 17 with command semantics and 13
when post hoc analysis adds metadata available at decision time. Across 11
fixed-SHA project executions, the same retrospective ladder reaches zero
metadata residuals; two exact mappings recur across three product frontends.

For prospective recovery, \emph{effect-bound records} commit frozen,
source-backed effect predictions and provenance before authorization. On 17
target-executed workflows from a prespecified holdout, predictions achieve 0.926
macro recall and 0.941 macro precision. Binding them cuts residual effect
instances from 10 to 3. The evaluated policy also requests another approval for
17/18 protocol cases, including cases whose realized traces add no effect class
beyond the command-only record; an exploratory offline replay shows that a
stricter added-effect ASK rule reduces that demand to 7/18 while leaving 2/7
command-residual cases without escalation. A Claude Code hook can place one frozen record
on an existing permission path. On that \texttt{npm run build} pair, a study
metadata gate does not launch the network arm after the metadata denial.
\end{abstract}

\section{Introduction}

An approval mechanism stores a durable record and later policy or audit can
consume that record. The record can name the approved command exactly while
omitting an effect of the workflow that command starts. We call this omission
\emph{\approval}: relative to an operational vocabulary, the durable record
faithfully identifies the entry invocation and leaves at least one effect of
its transitive workflow unbound.

A coverage gap is a representation property. It is an authorization failure
when a pre-specified effect policy requires different decisions for two
executions and their complete policy-visible inputs \(I(b)\) are identical.
Lemma~1 shows that no record-only policy can then guarantee both outcomes.
The same
transitive workflow can arise when a person runs the command directly; the
operational semantics of expansion do not depend on who typed the entry
string. What this paper measures is the decision input and durable record of
an independent approval mechanism that agents insert before that workflow
runs. Coding agents are therefore the measured setting
~\cite{codexapprovals,claudecodepermissions,copilotcliapprovals}.

Section~\ref{sec:coverage-gaps} uses two fixed-SHA cases as representation
examples. In Vite, an approved package-install command invoked
\texttt{postinstall} and wrote a workspace file. In an MCP case, a named
documentation-tool call exercised both its represented MCP capability and
network authority. Each record identified the entry invocation, and the
workflow crossed a wider operational boundary. These cases locate where
explicit fields, command semantics, and instance metadata diverge.

Documented npm, PEP~517, Cargo, Docker, Git, VS Code, and MCP semantics
expose the expansion points through which one invocation activates behavior
selected by project, dependency, or configuration artifacts
~\cite{npmscripts,pep517,cargobuildscripts,dockerfile,githooks,vscodetasks,
mcp2026tools}. Workflow expansion is the mechanism. Approval laundering is
the resulting record-coverage failure. Ordinary toolchain semantics produce
structural instances. An actor who controls an expansion source before the
decision can use the same channel to place a policy-relevant effect behind a
faithfully identified entry invocation.

\textbf{Research gap.}
Prior work addresses approval--execution binding, task-level closure, delegated
capabilities, canonical actions, tool admission, and user permissions
~\cite{consentintegrity,authorizationexecutiongap,chaincaps,authbench,
agentpermissionssurvey,pcaactions,cava,toolguardian}. Those lines constrain
which action executes, which authority a task requires, or whether the
displayed action is the action at the boundary. We measure a different
property of an already identified invocation: whether its decision-time
approval object covers the transitive operational effects of the selected
implementation, and when a missing effect changes a pre-specified
authorization decision.

\begin{figure*}[!t]
  \centering
  \includegraphics[width=\textwidth]{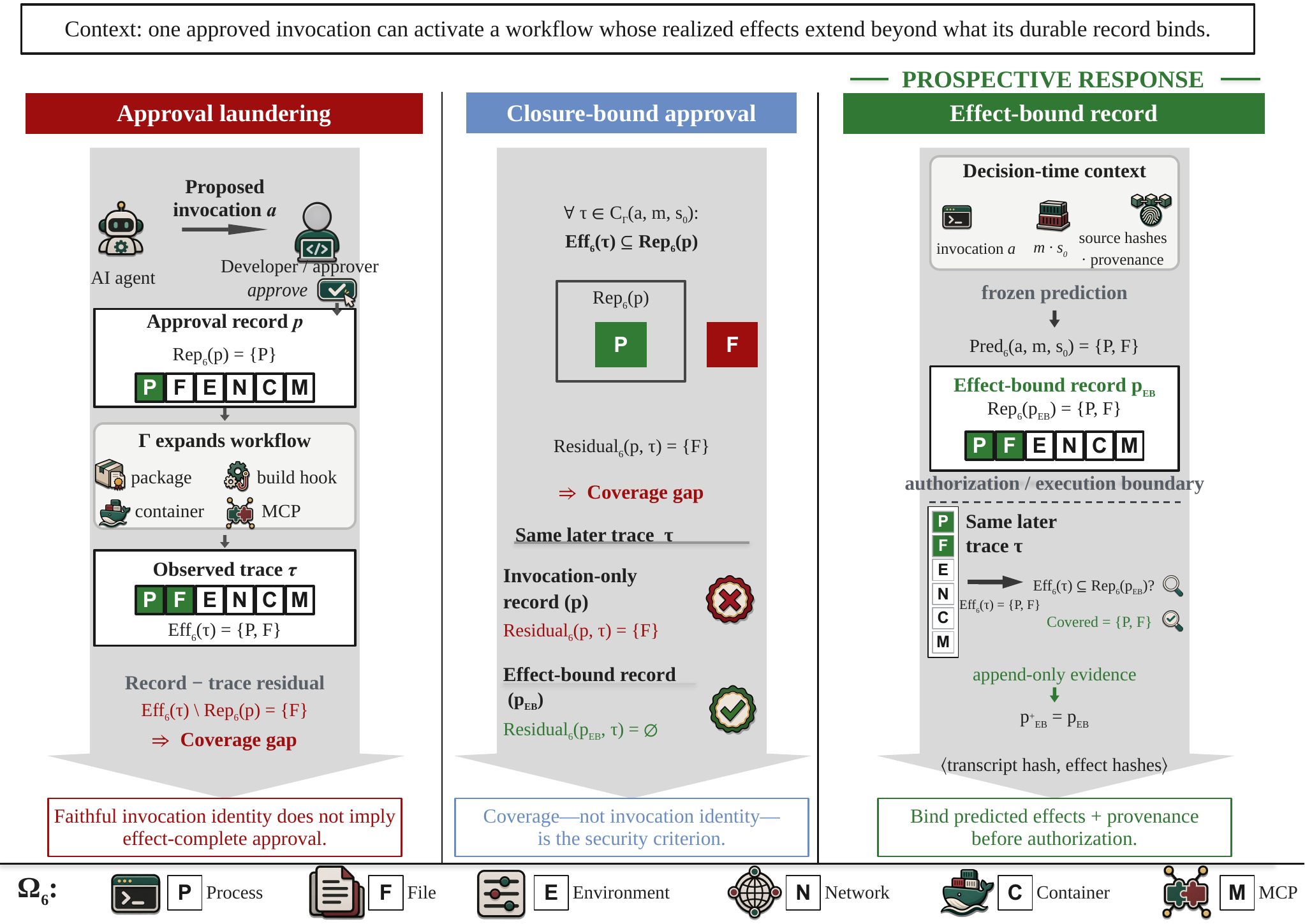}
  \caption{Approval laundering and its effect-bound response. A record can
  truthfully bind the entry invocation yet omit a transitive file effect (left);
  closure-bound approval marks criterion violation (cross) or satisfaction
  (check) beside the residual (center); an effect-bound record binds predictions
  and provenance before authorization and appends execution evidence afterward
  (right).}
  \label{fig:mechanism}
\end{figure*}

\textbf{Our approach.}
To make this gap measurable, we first formalize closure-bound approval over six
operational effect classes and derive a record-only indistinguishability result.
We then instantiate the property in the Approval-to-Action Security Benchmark,
which binds prompt-visible fields and inspectable decision-time metadata to
immutable execution evidence under three nested information views. Finally, we
evaluate prospective recovery from frozen decision-time metadata, bind the
resulting effect predictions and provenance into an effect-bound record before
authorization, and integrate that record into an approval-path prototype.
Figure~\ref{fig:mechanism} places the structural failure, its closure-bound
security condition, and the prospective effect-bound response side by side.

\textbf{Research questions.}
The problem decomposes into three questions:

\emph{RQ1: Approval-record coverage.} How does approval-record coverage change
across explicit-field, command-aware, and metadata-aware views, overall and by
source?

\emph{RQ2: Cross-project and cross-frontend recurrence.} Does the same
approval-to-effect gap recur in fixed-SHA public-project workflows and across
coding-agent frontends?

\emph{RQ3: Pre-execution recovery.} Can decision-time metadata predict and
bind missing effects before execution, and how often does the resulting policy
request additional approval?

\textbf{Results and contributions.}
This paper makes three contributions:
\begin{itemize}
    \item \emph{The record-to-closure property and its information limit.}
    We formalize \approval as a vocabulary-relative coverage gap and define
    closure-bound approval over six operational effect classes. A gap is a
    structural witness only after a permitting decision and a valid execution
    binding; it is an authorization failure only when the omitted effect
    changes a pre-specified decision. When the complete policy-visible inputs
    are identical and the required decisions differ, no deterministic or
    randomized record-only policy can guarantee both.

    \item \emph{Where coverage gaps occur.}
    We bind captured approval objects to append-only, hash-bound execution
    evidence across nine workflow mechanisms. Across 111 approval-object/trace
    pairs held fixed across views, records omitting at least one observed
    effect fall from 40 under explicit fields to 17 with command semantics and
    13 when post hoc analysis adds metadata available at decision time. Across
    11 evidence-complete executions of eight fixed-SHA cases, the count falls
    from 10 to 2 to 0. This ladder locates the gap.

    \item \emph{What decision-time metadata discloses, and what it costs.}
    On 17 target-executed holdout workflows, frozen predictions attain 0.926
    macro recall, 0.941 macro precision, and 13/17 exact effect sets, against
    9/17 for the command-only baseline. Binding those predictions raises
    case-macro disclosed-effect coverage from 0.755 to 0.922 and reduces
    residual effect instances from 10 to 3.     The evaluated policy routes 17/18
    protocol cases to an additional approval, including 9/17 reached cases
    whose realized traces add no effect class beyond the command-only record.
    A Claude Code hook
    carries one frozen record onto an existing permission path. The resulting
    noninteractive denial shows path compatibility.
\end{itemize}

\section{The Approval-Representation Ladder}
\label{sec:coverage-gaps}

Approval-time information forms a representation ladder. Explicit record fields
state what the interface binds directly. Generic command semantics recover
effects implied by the invocation class. Instance-specific metadata exposes
effects that the first two layers omit. The following cases isolate the two
successive transitions: Vite moves from explicit fields to command semantics,
and MCP moves from command semantics to decision-time metadata. Both are
representation examples.

\subsection{Explicit-Field Gap: Vite}

Consider a dependency installation in \path|vitejs/vite| at commit
\path|64dfee12| using the repository-pinned pnpm 10.34.3. An AI coding agent
requested approval for one command:

\begin{quote}
\small\ttfamily
pnpm --dir <fixed-sha-workspace> install --frozen-lockfile
\end{quote}

\begin{figure*}[t]
\centering
\includegraphics[width=\textwidth]{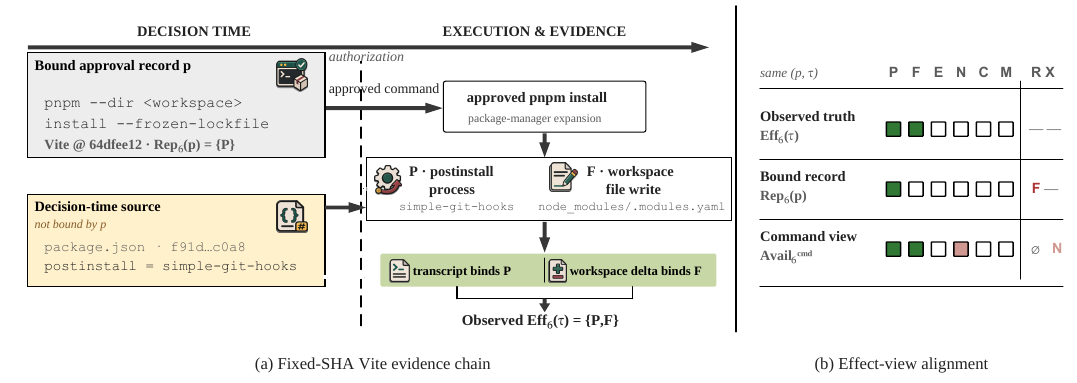}
\caption{Fixed-SHA Vite evidence chain and effect-view alignment for one
Claude Code run.
(a) Post-execution artifacts bind the realized \texttt{process} and
\texttt{file} effects of the approved install. (b) The study-encoded bound record misses \texttt{file}; the
conservative command view covers it but overpredicts \texttt{network}.}
\label{fig:vite-case}
\end{figure*}

\textbf{Approval-time representation.}
On one Claude Code fixed-SHA run of this install, the native approval surface
showed the exact \texttt{pnpm \ldots{} install --frozen-lockfile} command. The
study encodes that surface as a process-only representation. A driver log on the same run records a bash approval
choice and later turn completion; those events are driver-observed decision
evidence, not a product-emitted receipt object. The approved command was the
entry point to an installation workflow rather than the complete workflow that
executed. Figure~\ref{fig:vite-case} separates the approval-time expansion
source, the effects realized after authorization, and their alignment under the
bound-record and command views.

\textbf{Executed workflow.}
After that driver-observed approval, package-manager semantics invoked
\texttt{postinstall}. The same-run transcript binds
\texttt{simple-git-hooks}, and the workspace delta binds
\path|node_modules/.modules.yaml|. The study-encoded record therefore
represents a process effect, while the executed workflow exercises both process
and file effects.

\textbf{Coverage across views.}
The explicit record leaves the file effect uncovered. Conservative
package-install semantics recover it before execution but also predict network
for this execution (Figure~\ref{fig:vite-case}(b)), closing the residual without
yielding an exact set.

\subsection{Command-Residual Gap: MCP}

The MCP case identifies the next boundary: command semantics omit an
instance-specific network effect.

\textbf{Approval-time representation.}
At commit \path|7b1170d1| of \path|modelcontextprotocol/servers|, GitHub Copilot
CLI and Qwen Code each issued a successful, named documentation-tool call
through the fixed-SHA product configuration \path|.mcp.json|. Their records
represented the MCP capability; generic tool-call semantics likewise exposed
only that capability.

\textbf{Executed workflow.}
Post-execution evidence binds both the named MCP calls and concrete remote
responses. The workflow therefore uses the represented MCP capability and also
exercises network authority.

\textbf{Coverage across views.}
The network effect remains uncovered under both the explicit-field and
command-aware views. The missing information is nevertheless available before
execution: the 123-byte instance-specific \path|.mcp.json| identifies the
configured transport. This case therefore locates the next gap between generic
invocation semantics and inspectable workflow metadata.

Vite and MCP thus locate different boundaries in one representation ladder.
In Vite, command semantics recover an effect omitted by explicit fields; in MCP,
instance metadata recovers an effect omitted by generic command semantics. The
next section turns this ladder into a property of the sealed approval record and
its execution closure.

\section{Closure-Bound Approval and Threat Model}
\label{sec:model}

The representation ladder culminates in one binding question: whether the
approval record covers the operational effects reachable through the approved
workflow. We formalize that problem as a coverage relation over traces, derive
an observable gap predicate and a record-only policy indistinguishability
result, and instantiate both over the six effect classes used in our
experiments.

The Vite install in Section~\ref{sec:coverage-gaps} fixes the symbols used
below. The entry invocation \(a\) is the approved \texttt{pnpm install}
command. The expansion-source snapshot \(m\) contains the lifecycle metadata
and reachable scripts inspectable before the decision. The initial-state
snapshot \(s_0\) is the fixed-SHA workspace committed for that execution.
Toolchain semantics \(\Gamma\) are those of the pinned pnpm invocation. The
approval object \(b\) carries the machine-readable effect assertion
\(\{\texttt{process}\}\). If the completed path permits with outcome \(d\),
the record is \(p=\mathrm{Seal}(b,d)\). The realized trace \(\tau\) exercises
\texttt{process} and \texttt{file}. The same symbols apply to the MCP case,
with network and MCP labels in place of the file label.

\subsection{Approval Transaction and Execution Closure}

An approval transaction separates the object presented for a decision from
the record produced by that decision. The \emph{approval surface} is the
information made available to the decision-maker. A base context
$b_0\in\mathcal{B}_0$ is committed before any adapter response; it identifies
the requested entry invocation $a$ and any effect assertions supplied to the
decision path.

A staged mediator may then issue a routing outcome
$h\in\mathcal{H}$, such as \textsc{ASK}, pass-through, or local denial. The
\emph{bound decision context} $b=\langle b_0,h\rangle\in\mathcal{B}$ is the
exact machine-readable basis committed before final authorization; for an
unstaged mediator, $h=\bot$.

Because the bound basis may carry more information than a particular policy
consumes, we distinguish its policy-visible projection. For a concrete policy
consumer, $I(b)\in\mathcal{I}$ denotes the complete decision-time input on
which its output distribution may depend; any other state affecting that
distribution is included in $I(b)$ or fixed across compared instances.
Audit-only identifiers and provenance may remain in $b$ without entering
$I(b)$. Let $\Delta(\mathcal{D})$ denote the distributions over final outcomes.
A record-only policy is a possibly randomized rule
$F:\mathcal{I}\rightarrow\Delta(\mathcal{D})$, with deterministic rules
represented by point masses.

The completed permission path draws a final outcome $d\sim F(I(b))$ and seals
the durable pre-execution record $p=\mathrm{Seal}(b,d)$, with $B(p)=b$ and
$D(p)=d$. $B(p)$ is not a second notion of approval-time visible context.
We call $b$ the decision-time \emph{approval object} and $p$ its sealed
\emph{approval record}.
$\mathrm{Permit}(D(p))$ holds only when the completed approval path authorizes
execution; in particular, the routing outcome $h=\textsc{ASK}$ is not the final
authorization $d$. Post-execution evidence may be appended to form $p^+$, with
$B(p^+)=B(p)$ and $D(p^+)=D(p)$; it never becomes decision-time information.
Here, $\mathrm{Seal}$ denotes semantic immutability of the transaction basis
and outcome, not a cryptographic seal primitive. A product may expose $p$ as a
native audit entry; in our empirical instantiation, a hash-bound evidence
bundle preserves contemporaneous product artifacts and supports exactly the
fields those artifacts establish.

Execution need not stop at the entry invocation. Given a decision-time
initial-state snapshot $s_0$ committed for execution, a snapshot of inspectable
expansion sources $m$, and identified toolchain semantics $\Gamma$, the
\emph{execution closure}
$C_\Gamma(a,m,s_0)$ contains every reachable trace causally attributable to
$\Gamma$'s expansion of $a$. The boundary begins when the permitted invocation starts
and ends when its workflow terminates, becomes quiescent, or reaches another
independently mediated request. It includes effects produced before a failed
workflow terminates and by causally attributable descendants within that
boundary. A realized trace $\tau$ is the \emph{executed object}.

We write $\mathrm{CtxBind}(p;a,m,\Gamma,s_0)$ when the pre-execution binding
commits the exact invocation, source contents and digests, toolchain identity,
and initial-state snapshot to the same transaction as $p$. For an observed
trace, $\mathrm{Bind}(p,\tau;a,m,\Gamma,s_0)$ additionally requires causal and
temporal attribution of $\tau$ to that committed context. An execution using
different committed content or identity requires a new decision context.
Nondeterministic behavior reachable from the fixed basis remains inside the
closure.

To compare an approval object with the executed object, let $\Omega$ be a
chosen vocabulary of coarse operational labels. Let
$\mathrm{Rep}_{\Omega}(b)\subseteq\Omega$ contain the machine-readable effect
assertions carried by the frozen basis $b$, and let $\alpha_{\Omega}$ project
the raw effects $\mathrm{Eff}(\tau)$ of trace $\tau$ into that vocabulary. For
a sealed record, we use the shorthand
$\mathrm{Rep}_{\Omega}(p):=\mathrm{Rep}_{\Omega}(B(p))$. We write
$\mathrm{Eff}_{\Omega}(\tau)=
\alpha_{\Omega}(\mathrm{Eff}(\tau))$.

\subsection{Threat Model}

\textbf{Actors and authorization boundary.}
We study one approval-mediated execution initiated by an AI coding agent. The
agent proposes $a$; a human principal or automated policy decides through the
approval mediator; and, when the completed path permits execution, the
developer toolchain expands $a$ into $\tau$. Although mediation originates in a
local coding-agent session, the resulting closure may cross process, file,
environment, network, container, and MCP boundaries before the next independent
mediation point.

\textbf{Structural and adversarial regimes.}
The same record--trace coverage failure arises under
\emph{structural occurrence}, when ordinary nonmalicious expansion exercises
an omitted effect label, and under \emph{adversarial exploitation}, when an
actor controls or influences an expansion source consumed by $\Gamma$ to
induce a policy-relevant omitted label behind the same faithfully identified
entry invocation. The regimes differ in source control, not in the coverage
predicate below.

\textbf{Adversary capabilities.}
Before $b_0$ is committed, the adversary may control or influence an
expansion-bearing workspace artifact, local tool configuration or state, or a
referenced implementation, for example through a repository contribution or a
dependency/supply-chain position. In the studied workflows, these sources
include package lifecycle definitions and
reachable scripts, build-backend selection and enabled build scripts,
Dockerfile instructions, configured Git hooks, named IDE tasks and their
reachable scripts, import-time code in discovered test modules, and MCP
configuration selecting a local command or remote transport. The adversary may
know the approval schema, $a$, and ordinary toolchain semantics, and choose the
source so that execution activates a target effect class. Normal toolchain
expansion carries the effect while the recorded entry invocation remains
unchanged.

\textbf{Trust boundary.}
The capabilities above stop at expansion sources chosen before \(b_0\) is
committed. The mediator is trusted for a narrower function. After commitment, \(p\) names
the invocation that was actually launched, \(b\) remains the basis of that
decision, and the temporal order of proposal, decision, and launch is
preserved. \(\mathrm{Seal}(b,d)\) is this semantic immutability of the
committed basis and outcome. It is a transaction requirement on the mediator.
The hash shows that the
preserved bytes were not altered after capture. Correspondence between the
approved content and the executed workflow holds when
\(\mathrm{CtxBind}\) and \(\mathrm{Bind}\) are satisfied for that run. A
stored digest does not create that correspondence by itself.

The measurement trusted computing base, used only to establish an empirical
witness, contains approval capture, record--trace association, artifact
hashing, effect collectors, and frozen projection rules. Executed workflow
code is assumed unable to forge those externally preserved objects. This
assumption is independent of the adversary's control of expansion sources:
the adversary chooses what the toolchain will expand, and the measurement
path records the mediator's decision and the resulting trace from outside
that workflow.

\textbf{Security objective.}
Three layers stay separate. A \emph{coverage gap} means an executed label is
absent from the record. A \emph{structural witness} is a coverage gap for
which the completed path permits execution and the trace is validly bound to
the approved context. An \emph{authorization failure} further requires a
pre-specified effect policy \(Q\) whose required decision changes when the
omitted label is present. A structural witness falsifies closure-bound
approval. It becomes an authorization failure only under that policy
condition.

Let \(\mathrm{AdvCtrl}(\omega,\tau;a,m,\Gamma,s_0)\) mean that adversarial
control of an expansion source causally induces label \(\omega\) in \(\tau\).
Adversarial success occurs when the completed path permits, the trace is
validly bound, and some omitted label
\(\omega\in\mathrm{Eff}_{\Omega}(\tau)\setminus\mathrm{Rep}_{\Omega}(p)\) is
induced by that control and changes the decision required by \(Q\). The
running policy is: \(Q\) permits an \texttt{npm install} whose closure stays
in \(\{\texttt{process}\}\), and \(Q\) re-mediates an attacker-controlled
\texttt{postinstall} that adds \(\texttt{file}\) behind the same
policy-visible input. If \(I(b)\) is identical for both executions, every
record-only rule \(F\) produces the same decision distribution and cannot
guarantee both required outcomes. Lemma~1 states that information limit. If
\(I(b)\) contains the differing script text or another field the policy can
read, the premise \(I(B(p_1))=I(B(p_2))\) fails and the lemma does not apply.
Removing the adversarial-control condition yields the corresponding
structural witness.

\textbf{Practicality of the threat model.}
Documented developer toolchains expose executable expansion points in project,
dependency, and configuration artifacts
~\cite{npmscripts,pnpmbuildsettings,pep517,cargobuildscripts,dockerfile,
githooks,vscodetasks}. Under pnpm 10, dependency install scripts are eligible to
execute only when permitted by project build settings
~\cite{pnpmbuildsettings}. Within the source-control assumptions above, the
adversary controls an expansion source that the selected toolchain already
consumes. In the MCP case, committed
configuration selects the transport and server endpoint behind the recorded
tool name.

\subsection{Closure-Bound Approval and Witnessed Violation}

For those closures, the relevant criterion is coverage rather than object
equality. A record may conservatively represent more effects than one execution
realizes; the failure
of interest occurs when an executed effect is absent from the represented set.

\textbf{Definition 1 (Closure-bound approval).}
For a fixed transaction context $(p,a,m,\Gamma,s_0)$ satisfying
$\mathrm{Permit}(D(p))$ and $\mathrm{CtxBind}(p;a,m,\Gamma,s_0)$, an approval is
closure-bound with respect to $\Omega$ when
\begin{equation}
 \forall \tau\in C_\Gamma(a,m,s_0):
 \mathrm{Eff}_{\Omega}(\tau)\subseteq\mathrm{Rep}_{\Omega}(p).
 \label{eq:closure-bound}
\end{equation}
Conservative overrepresentation satisfies this one-sided relation, whereas one
omitted label does not. A command-only record may therefore identify the entry
invocation exactly while
leaving effects triggered by its transitive workflow unstated.

\textbf{Definition 2 (Approval-object coverage gap).}
For a captured approval object $b$ and an observed trace
$\tau\in C_\Gamma(a,m,s_0)$, define the pure coverage predicate
\begin{equation}
 \mathrm{Gap}_{\Omega}(b,\tau)\iff
 \mathrm{Eff}_{\Omega}(\tau)\not\subseteq\mathrm{Rep}_{\Omega}(b).
 \label{eq:coverage-gap}
\end{equation}
For a completed record $p=\mathrm{Seal}(b,d)$, a pair $(p,\tau)$ is a measured
\emph{structural approval-laundering witness} when the completed path permits,
$\tau$ is validly bound to $(p,a,m,\Gamma,s_0)$, and
$\mathrm{Gap}_{\Omega}(B(p),\tau)$ holds. It is adversarial when an omitted
label $\omega$ is also causally induced by adversarial control over an
expansion source.

\textbf{Observation 1 (One-sided falsification).}
One structural witness decisively falsifies the closure-bound condition
in~\eqref{eq:closure-bound} for its fixed $(p,a,m,\Gamma,s_0,\Omega)$ context.
The empirical test is one-sided: observed witnesses falsify the universal
property; an observed run without a witness does not establish the property
over the entire reachable closure.

The premise is equality of the complete policy-visible input
\(I(b)\), including every field the policy's output distribution may depend
on, and not equality of the command string alone.

\textbf{Lemma 1 (Record-only policy indistinguishability).}
Let $F:\mathcal{I}\rightarrow\Delta(\mathcal{D})$ be a possibly randomized
record-only decision rule, with deterministic rules represented by point
masses, and let $Q:\mathcal{P}(\Omega)\rightarrow\mathcal{D}$ be deterministic.
For an execution instance $i$, write
$q_i=Q(\mathrm{Eff}_{\Omega}(\tau_i))$. For two validly bound execution
instances, record-level indistinguishability with distinct required outcomes gives
\begin{equation}
 \begin{gathered}
 I(B(p_1))=I(B(p_2)),\qquad q_1\ne q_2,
 \\[-1pt]
 \Longrightarrow\quad
 F(I(B(p_1)))=F(I(B(p_2)))=:\mu,
 \end{gathered}
 \label{eq:record-only-indistinguishability}
\end{equation}
and $F$ cannot agree with $Q$ almost surely on both instances: it is impossible
that $\mu(\{q_1\})=\mu(\{q_2\})=1$.
Appendix~\ref{app:lemma-proof} gives the proof.

This is the information requirement behind the representation ladder. Because
$I(B(p))$ may include the exact invocation, $F$ may interpret generic command
semantics. Whenever policy-relevant executions remain indistinguishable under
that projection, any record-only policy that implements $Q$ with probability
one requires additional decision-time information that separates them. If no
such information is available, the decision path must conservatively represent
the union, re-mediate, reject, or contain the unresolved effects. Randomization
can trade off error across the two instances, but it cannot recover the missing
distinction or guarantee $Q$ on both.

\subsection{Six-Class Operationalization}

The experiments instantiate $\Omega$ as the six-class operational vocabulary
$\Omega_6$: process, file, environment, network, container, and MCP. The
environment class denotes mutation of session or persistent environment state,
and MCP denotes a client-observed protocol capability or successful named tool
call~\cite{mcp2025tools,mcp2025lifecycle,mcp2026tools,mcp2026versioning}.

The six labels deliberately span four operational dimensions: process, file,
and network denote operations; environment denotes a mutation; container
denotes a resource boundary; and MCP denotes a protocol capability. Together,
they measure operational effect-class coverage at this abstraction.
Destination, intent, severity, credentials, path identity, and
product-contract status remain separate policy dimensions outside
\(\Omega_6\). A policy that forbids every network effect can be evaluated from
the network label. A policy that allows build-output writes and forbids access
to one sensitive path needs the path and the operation; the file label does
not express that distinction. Coverage is the representation-layer
completeness condition. RQ3 pairs it with precision and the rate of additional
approval requests so that conservative disclosure is measured against
selectivity.

Accordingly, $\mathrm{Eff}_6(\tau)$ abbreviates
$\alpha_6(\mathrm{Eff}(\tau))$, and $\mathrm{Rep}_6(b)$ contains only labels
carried by machine-readable effect assertions in the frozen approval object.
For a completed record, $\mathrm{Rep}_6(p)=\mathrm{Rep}_6(B(p))$.

Figure~\ref{fig:vite-case} illustrates the measurable predicate. Its explicit
projection is $\mathrm{Rep}_6(b)=\{\texttt{process}\}$, whereas the bound trace
yields $\mathrm{Eff}_6(\tau)=\{\texttt{process},\texttt{file}\}$. Therefore
$\mathrm{Gap}_6(b,\tau)$ holds; if the completed path permits and the trace is
validly bound, the same pair is a structural witness.

Mechanism classes describe how an invocation
expands, whereas effect labels describe the resulting operation, mutation,
resource boundary, or protocol capability. The benchmark carries this
comparison forward:
Section~\ref{sec:benchmark} fixes the captured approval basis and inspectable
sources before execution and derives labels only from post-execution evidence.
Where the completed decision and full transaction binding are observed, it
also instantiates the permission-bearing record required for a witness.

To measure decision-time recoverability for a fixed approval basis, let
$\mathrm{Avail}^{v}_{6}(b,a,m,\Gamma,s_0)$ denote the labels recoverable under
information view
$v\in\{\mathrm{explicit},\mathrm{command},\mathrm{metadata}\}$.
Section~\ref{sec:benchmark} evaluates these fixed views.
Section~\ref{sec:audit} then makes the metadata prediction a durable
approval-time assertion by committing it to $b$ before the decision and
preserving it in $p$.

\section{The Approval-to-Action Security Benchmark}
\label{sec:benchmark}

The Approval-to-Action Security Benchmark fixes captured approval fields and
the inspectable source snapshot before execution, associates them with
immutable post-execution evidence, and evaluates the Definition~2 coverage gap
on each approval-object/trace pair. Over the same fixed pair, it compares the explicit-field,
command-aware, and metadata-aware $\mathrm{Avail}^{v}_{6}$ views. The
explicit-field view tests the captured approval basis; the richer views estimate
counterfactual decision-time recoverability.

\subsection{Benchmark Overview and Temporal Pipeline}

The pipeline follows the order approval object, completed decision, execution,
and evidence. For captured RQ1 instances, the benchmark preserves the approval
object $b$, identified invocation $a$, source snapshot $m$, and initial-state
snapshot $s_0$. When contemporaneous product artifacts also establish final
outcome $d$, the same transaction instantiates $p=\mathrm{Seal}(b,d)$;
Table~\ref{tab:formal-empirical} specifies the corresponding claim
eligibility. The prospective pipeline constructs a distinct approval object by
freezing a prediction and its source/rule hashes. The policy evaluation
instantiates routing outcome $h$ over that object, while the live hook
separately observes the handoff from $h$ to a final product decision;
Section~\ref{sec:audit} gives the construction. When that path permits
execution, identified toolchain semantics $\Gamma$ expand $a$ and realize one trace
$\tau\in C_\Gamma(a,m,s_0)$. Only afterward do mechanism-specific collectors
preserve transcripts, state deltas, and hashes from which
$\mathrm{Eff}_6(\tau)$ is derived.

These paths support two no-hindsight checks: RQ1 compares
$\mathrm{Eff}_6(\tau)$ with $\mathrm{Rep}_6(b)$, whereas RQ3 compares it with
the frozen $\mathrm{Pred}_6(a,m,s_0)$.

\begin{figure*}[!t]
\centering
\includegraphics[width=\textwidth]{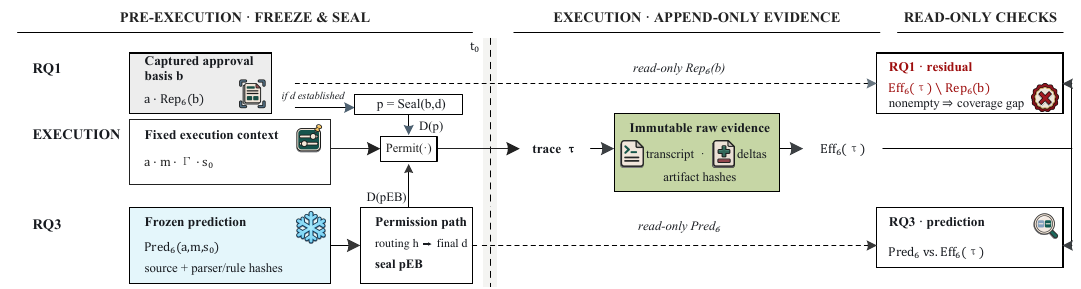}
\caption{Forward-only benchmark protocol. RQ1 freezes the captured approval
basis $b$, while RQ3 freezes a prediction before $t_0$; their decision inputs
converge with the fixed execution context at a shared permission and evidence
spine. The resulting effect projection feeds both read-only checks. Where
contemporaneous decision evidence establishes $d$, the completed record
$p=\mathrm{Seal}(b,d)$ feeds $D(p)$ to $\mathrm{Permit}(\cdot)$. Appended
hashes leave the decision-time state unchanged. Coverage is scored on $b$;
Table~\ref{tab:formal-empirical} operationalizes the additional evidence
required for a permission-bearing witness.}
\label{fig:benchmark-pipeline}
\end{figure*}

Appendix Figure~\ref{fig:benchmark-pipeline} shows their shared forward-only
evidence discipline; post-execution evidence enters scoring but never the
approval basis or recorded prediction.

\textbf{Protocol invariants.}
Across this pipeline, every approval-time field used by a view, the invocation,
captured source contents and identities, identified toolchain semantics, and
initial-state snapshot are fixed before the execution they are used to
evaluate; evidence attribution must preserve their approval-object--trace
binding.

\noindent\textbf{Formal-to-empirical correspondence.}
For run $r$, the hash-bound bundle $\mathcal{E}_r$ preserves the available
approval surface, decision evidence, execution context, and trace evidence.
Table~\ref{tab:formal-empirical} maps these artifacts to the logical objects
and their acceptance tests.

\begin{table*}[t]
\caption{Formal-to-empirical correspondence. For each run $r$, the hash-bound
bundle $\mathcal{E}_r$ preserves the recorded evidence used to instantiate each
logical object; the right column gives the corresponding acceptance condition.}
\label{tab:formal-empirical}
\centering
\footnotesize
\setlength{\tabcolsep}{3pt}
\rowcolors{2}{white}{black!4}
\begin{tabular}{@{}>{\raggedright\arraybackslash}p{0.09\textwidth}
>{\raggedright\arraybackslash}p{0.57\textwidth}
>{\raggedright\arraybackslash}p{0.27\textwidth}@{}}
\toprule
\rowcolor{white}
Object & Recorded evidence & Acceptance condition \\
\midrule
$a,b_0$ & Exact invocation; captured prompt or fixture approval surface and its
displayed effect fields & Capture precedes the adapter/final decision; surface
and launched invocation match; capture hash verifies \\
$h,b$ & Adapter routing outcome, or $h=\bot$ when no routing stage exists;
bound basis $b=\langle b_0,h\rangle$ & Same transaction; $b$ is committed before
final authorization \\
$d,p$ & Native decision receipt or transcript/driver event for the exact
invocation; sealed basis and outcome & \textsc{ASK} is not permission; $d$
precedes launch; missing $d$ remains unobserved \\
$m,\Gamma,s_0$ & Source contents and digests or fixed SHA; tool, agent, and
version; pre-run workspace/configuration state & Identities and hashes belong to
run $r$ \\
$\tau,\mathrm{Bind}$ & Non-prompt transcript, event log, state delta, and
mechanism-specific effect evidence & Same run; temporal and causal attribution;
artifact hashes verify \\
\bottomrule
\end{tabular}
\end{table*}

\noindent\textbf{Claim eligibility.}
These encodings define nested claim gates. A pair is \emph{coverage-complete}
when $\mathcal{E}_r$ preserves $b$ and an attributable $\tau$; all 111 canonical
RQ1 pairs satisfy this gate. It is \emph{decision-complete} only when a receipt
or transcript/driver event records the completed ALLOW or DENY for the exact
invocation, and \emph{bind-complete} only when the context and ordering tests in
Table~\ref{tab:formal-empirical} establish
$\mathrm{Bind}(p,\tau;a,m,\Gamma,s_0)$. Coverage completeness supports
$\mathrm{Gap}_6(b,\tau)$; only decision and binding completeness together with
$\mathrm{Permit}(D(p))$ support a structural witness. Thus calibration,
product capture, and permission-bearing evidence retain distinct roles.

\noindent\textbf{Structural-witness eligibility on the 40 explicit gaps.}
We audited every explicit coverage gap under a frozen codebook that marks each
of $\mathrm{Permit}$, pre-authorization context binding, and post-execution
trace attribution as \texttt{proven}, \texttt{contradicted}, or
\texttt{insufficient}. Import timestamps are not authorization times, and a
repository SHA alone is not $\mathrm{CtxBind}$. The 111 canonical records
contain 40 explicit-field coverage gaps. Of those 40, 8 have driver-observed
permitting evidence, and 0 pass full context binding and trace attribution, so
the structural-witness count is 0.
Appendix Table~\ref{tab:witness-funnel} reports the stratum funnel.
Coverage gaps and structural witnesses therefore remain distinct empirical
claims.

\subsection{Unit of Analysis and Evidence Strata}

Every execution attempt contributes one append-only ledger row with an
immutable terminal snapshot. The ledger retains 149 attempts: 139 terminate in
\path|SUCCESS|, while 10 retain a non-success terminal status. Non-success
attempts remain part of the ledger but do not enter the primary RQ1 denominator.

Retries must not receive additional analytical weight. At the same time, a
corrected capture must be able to supersede an earlier successful snapshot. The
canonical RQ1 view therefore retains the latest successful snapshot for each
\emph{agent--task} key, yielding 111 \emph{deduplicated successful agent--task
records}; retries thus contribute at most one successful record per key. Each
canonical record is paired with its observed trace. RQ1 scores Definition~2
coverage gaps on the 111 coverage-complete pairs;
Table~\ref{tab:formal-empirical} governs the additional eligibility for a
structural witness. A qualifying structural witness therefore falsifies
closure-bound approval for its fixed context under Observation~1; adversarial
success additionally requires $\mathrm{AdvCtrl}$.
Appendix Table~\ref{tab:evidence-state} gives the complete terminal-state and
deduplication accounting for this denominator.

These 111 approval-object/trace pairs form the RQ1 denominator; source provenance
determines which claims each pair can support. The \emph{harness/parser fixture stratum}
isolates npm lifecycle scripts, PEP~517 hooks, Cargo build scripts, git hooks,
IDE tasks, nested shells, test imports, Docker builds, and MCP configuration
under known conditions. The \emph{cross-frontend construct control} uses the
same fixtures across frontends and requires a captured approval prompt, a
non-prompt transcript or driver artifact, agent metadata, observed effects, and
hash-bound copied artifacts.
\emph{Real-project executions} bind protocol-selected fixed-SHA public
repositories to concrete post-approval evidence. \emph{Prospective holdouts}
instead bind parser predictions over unmodified raw workflow metadata to
outcomes observed only after those predictions are frozen.

These strata have distinct analytical roles. The full RQ1 corpus contains nine
successful mechanism classes. Harness/parser fixtures isolate the coverage
construct, the shared-fixture stratum controls frontend capture, and qualifying
product and fixed-SHA evidence can additionally pass the witness-eligibility
gate above. RQ2 interprets fixed-SHA real-project executions
alongside that cross-frontend construct control, which
covers eight classes; Docker is represented in the harness/parser fixture and
fixed-SHA project strata. RQ3 keeps its controlled, natural-project development,
and confirmatory fixed-SHA cohorts separate from the retrospective views.

\subsection{RQ1: Approval-Record Coverage}

RQ1 holds the 111 coverage-complete approval-object/trace pairs fixed and asks
how coverage changes as decision-time information is enriched.
On the approval side, let $\mathrm{Cmd}_6(a;\Gamma)$ denote conservative generic
semantics of exact invocation $a$ under the identified toolchain. The
metadata-aware view uses parser v1.1, developed from the controlled cohort and a
natural-project development cohort (A), then frozen before a disjoint fixed-SHA
confirmatory cohort (B); Section~\ref{sec:audit} gives the full
development-and-freeze sequence. Its output
$\mathrm{Pred}_6(a,m,s_0)$ is the full pre-execution set emitted by the frozen
predictor for the fixed
identified $\Gamma$ (suppressed in its argument list): it includes the
command/surface baseline and labels inferred from transitive metadata, so
$\mathrm{Cmd}_6(a;\Gamma)\subseteq\mathrm{Pred}_6(a,m,s_0)$ in the evaluated
pipeline. Suppressing common arguments, the three available sets are
\begin{equation}
\begin{aligned}
\mathrm{Avail}^{\mathrm{explicit}}_6
  &= \mathrm{Rep}_6(b),\\
\mathrm{Avail}^{\mathrm{command}}_6
  &= \mathrm{Rep}_6(b)\cup\mathrm{Cmd}_6(a;\Gamma),\\
\mathrm{Avail}^{\mathrm{metadata}}_6
  &= \mathrm{Rep}_6(b)\cup\mathrm{Pred}_6(a,m,s_0).
\end{aligned}
\label{eq:available-sets}
\end{equation}
On the trace side, mechanism-specific collectors bind the observed process (P),
file (F), environment (E), network (N), container (C), and MCP (M) classes to
raw post-execution artifacts. The resulting operational projection spans
operations, state mutation, resource boundaries, and protocol capability, with
every label grounded in post-execution evidence.

The frozen codebook requires an environment effect to mutate session or
persistent environment state. A child workflow that only reads an inherited
canary, or declares a task-local variable, does not contribute an environment
effect. The MCP class requires either client-observed capability availability
under the effective protocol revision, established through legacy
initialization or modern discovery and per-request metadata, or a named
successful tool call.
Fixture-authored protocol-shaped output alone is excluded
~\cite{mcp2025tools,mcp2025lifecycle,mcp2026tools,mcp2026versioning}.
Accordingly, an approved configuration that reports \path|MCP_NOT_VISIBLE| and
has no post-approval child-process evidence contributes no strict M or P effect.

Under the view-specific sets in~\eqref{eq:available-sets}, a pair remains
residual under view $v$ when
$\mathrm{Eff}_6(\tau)\not\subseteq
\mathrm{Avail}^{v}_6(b,a,m,\Gamma,s_0)$. By~\eqref{eq:coverage-gap}, the
explicit residuals are observed Definition~2 coverage gaps. Reductions under
the richer views estimate
counterfactual decision-time recoverability while the captured approval object
$b$ remains unchanged.
Appendix~\ref{app:supplementary-evidence} reports the trace-preserving
reclassifications and the first-success sensitivities of the explicit, command,
and metadata views.

\noindent\textbf{Results.}

\finding{1}{Explicit approval fields omit observed effect classes in every
benchmark mechanism class.}{Under the captured explicit-field view, 40 of the
111 canonical records have a nonempty
$\mathrm{Eff}_6(\tau)\setminus\mathrm{Rep}_6(b)$ residual. Every one of the
nine mechanism classes contributes at least one gap, establishing the
occurrence of explicit-field coverage failures throughout this benchmark.}

Appendix Table~\ref{tab:rq1-mechanism-effects} localizes those 40
explicit-field gaps and tracks the remaining residual records in each
mechanism class after command and metadata information is added.

We next apply the command-semantic and frozen-metadata enrichments to the
explicit-field gaps. The frozen metadata predictions are computed over 61
hash-bound source files. Neither enrichment consumes expected-effect annotations
or post-execution labels.

For comparison, an annotation-backed reference contrasts prompt-only inference
with replay of explicit expected-effect annotations over the same 111 tasks.
Appendix~\ref{app:supplementary-evidence} reports this comparison alongside the frozen
non-annotation views.

\finding{2}{Command semantics resolve 23 of 40 residual records; metadata
resolves four more.}{Across the same 111 records, residual records fall from 40
under explicit fields to 17 after generic command semantics and to 13 after raw
metadata. All 13 remaining residual records belong to the harness/parser
fixture stratum.}

Within that gradient, the explicit, command, and metadata residual counts are
30/15/13 on the 60 harness/parser fixtures, 0/0/0 on the 40 cross-agent
fixtures, and 10/2/0 on the 11 fixed-SHA executions. This same-corpus split
isolates the contribution of each information layer. The next section first
tests recurrence across fixed-SHA projects and product frontends.

\section{Fixed-SHA Recurrence Across Projects and Coding-Agent Frontends}
\label{sec:recurrence}

RQ2 asks whether the decision-time representation gradient persists in
fixed-SHA public projects and recurs across product frontends.

We test those two steps in order: first individual executions in a fixed-SHA
project stratum, then exact represented-to-observed mappings across frontends
at the same repository revision.

\begin{figure}[!ht]
\centering
\includegraphics[width=\columnwidth]{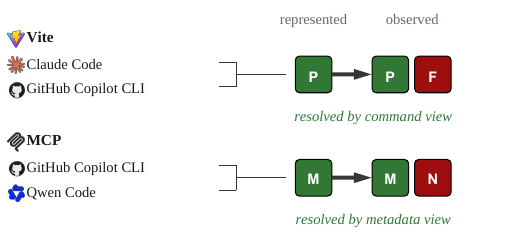}
\caption{Exact mappings recur across product frontends; a construct-control
view tests capture comparability. A strict label projection over the same 40
shared-fixture records retains all 40 and yields $0/40$ strict gaps
after projecting 36 inherited-read E and four protocol-shaped M labels.}
\label{fig:exact-mapping-construct-control}
\end{figure}

The paired executions remain nested within the 11 fixed-SHA executions and the
enclosing 111-record corpus. A separate cross-frontend construct control uses
shared fixtures to test whether frontend-specific capture behavior could create
apparent recurrence outside those paired executions.
Figure~\ref{fig:exact-mapping-construct-control}
shows the exact cross-frontend mappings and the construct-control view.

\subsection{Fixed-SHA Workflow Recurrence}
\label{sec:real-projects}

To test whether RQ1's approval-record coverage gradient persists in real
dependency and build graphs, we construct a fixed-SHA project stratum spanning
npm lifecycle, PEP~517, Cargo, Docker, and MCP workflows. The
analysis extends the comparison to public-repository executions with file,
container, and network effects.

\noindent\textbf{Outcome-blind case selection.}

Before any execution entered the ledger, a deterministic fixed-SHA scan of
nine predefined public repositories produced 81 mechanism-bearing metadata
findings. From those findings, the protocol fixed eight
repository--mechanism cases using repository provenance, source hashes, and
supported mechanisms.

Those eight fixed-SHA repository--mechanism cases yield 11 imported executions:
six cases contribute one execution, Vite two, and
\path|modelcontextprotocol/servers| three. Because gap status is computed
from each run's approval object and observed effects, the analysis evaluates
these 11 executions individually.

\Needspace{4\baselineskip}
\noindent\textbf{Evidence protocol.}

Each evidence-complete execution contains a captured approval record, agent
metadata, a non-prompt transcript or driver artifact, observed-effect evidence,
and a hash-bound snapshot. The append-only ledger retains capture attempts;
canonicalization fixes the 11 evidence-complete imported executions before
coverage-gap computation. Pre-execution metadata identifies workflow expansion,
while artifacts captured after approval bind the observed effects.

This three-view comparison yields the following fixed-SHA result.

\finding{3}{Residual records fall from 10 to 2 to 0 across fixed-SHA
workflows.}{Across 11 evidence-complete executions of eight
repository--mechanism cases, explicit fields leave 10 residual records,
command semantics leave 2, and post hoc metadata leaves none.}

All eight cases contribute at least one explicit-field gap, and the two
command-aware residual records are successful MCP calls.
Table~\ref{tab:real-projects} reports all three residual views and the
corresponding expansion source per case; Appendix
Table~\ref{tab:fixed-sha-evidence} binds each case to its post-approval
evidence.

\begin{table*}[!t]
\caption{Per-case residual records across 11 fixed-SHA runs. Explicit fields,
command semantics, and post hoc metadata leave 10, 2, and 0 residual records,
respectively. P, F, C, N, and M denote process, file, container, network, and
MCP.}
\label{tab:real-projects}
\centering
\footnotesize
\setlength{\tabcolsep}{2pt}
\rowcolors{3}{black!4}{white}
\begin{tabular}{@{}>{\raggedright\arraybackslash}p{0.19\textwidth}cccc
>{\raggedright\arraybackslash}p{0.24\textwidth}
>{\raggedright\arraybackslash}p{0.13\textwidth}@{}}
\toprule
Case and short SHA & Runs & \multicolumn{3}{c}{Residual records} &
Mechanism / source & Represented $\rightarrow$ observed \\
\cmidrule(lr){3-5}
& & Explicit & Command & Metadata & & \\
\midrule
\path|webpack/webpack| \texttt{@efeb5ce2} & 1 & 1 & 0 & 0 &
npm lifecycle / \path|package.json| & P $\rightarrow$ P,F \\
\path|vitejs/vite| \texttt{@64dfee12} & 2 & 2 & 0 & 0 &
npm lifecycle / \path|package.json| & P $\rightarrow$ P,F \\
\path|pypa/pip| \texttt{@72e6c594} & 1 & 1 & 0 & 0 &
PEP~517 / \path|pyproject.toml| & P $\rightarrow$ P,F \\
\path|pallets/flask| \texttt{@36e4a824} & 1 & 1 & 0 & 0 &
PEP~517 / \path|examples/celery/pyproject.toml| & P $\rightarrow$ P,F \\
\path|BurntSushi/ripgrep| \texttt{@dfe4a81d} & 1 & 1 & 0 & 0 &
Cargo build script / \path|build.rs| & P $\rightarrow$ P,F \\
\path|microsoft/vscode| \texttt{@8d5908d6} & 1 & 1 & 0 & 0 &
Docker build / \path|.devcontainer/Dockerfile| & P $\rightarrow$ P,F,C \\
\path|docker/awesome-compose| \texttt{@30f4b7f6} & 1 & 1 & 0 & 0 &
Docker build / \path|angular/angular/Dockerfile| & P $\rightarrow$ P,F,C \\
modelcontextprotocol/\allowbreak servers \texttt{@7b1170d1} & 1 & 0 & 0 & 0 &
MCP client configuration / \path|.mcp.json| & M $\rightarrow$ $\emptyset$ (strict) \\
\emph{same fixed-SHA case} & 2 & 2 & 2 & 0 &
MCP remote calls / client \path|.mcp.json| & M $\rightarrow$ M,N \\
\midrule
\rowcolor{white}
\textbf{Total} & \textbf{11} & \textbf{10} & \textbf{2} & \textbf{0} &
\multicolumn{2}{l@{}}{\textbf{8 fixed-SHA cases across five mechanism families}} \\
\bottomrule
\end{tabular}
\end{table*}

\noindent\textbf{Evidence behind the gradient.}
Package, build, and container cases realize P$\rightarrow$P,F or
P$\rightarrow$P,F,C, and conservative command semantics cover every such row.
The two successful remote MCP executions realize M$\rightarrow$M,N, remain
command-residual, and are covered once instance-specific metadata is added; the
retained configuration attempt is the sole fixed-SHA non-gap because it
activates no capability. The next analysis tests
frontend-specific capture with the shared-fixture construct control.

\subsection{Cross-Frontend Exact-Mapping Recurrence}

\noindent\textbf{Capture comparability.}
The prespecified 42-cell frontend--fixture matrix yields 40 evidence-complete
bundles and two fail-closed refusals. The strict label projection places
inherited canary reads and fixture-authored protocol-shaped events outside the
coverage-gap construct while retaining all 40 records, yielding 0/40 strict
gaps across five frontends. This negative control rules out
capture-created strict gaps across the tested shared fixtures. Appendix
Table~\ref{tab:agent-configurations} reports the full configuration matrix; the
fixed-SHA paired analysis below tests exact-mapping recurrence.

To isolate frontend variation, we compare codebook-valid executions of the same
fixed-SHA case under distinct product frontends. The comparison spans two
qualitatively different expansion settings in the fixed-SHA stratum: Vite's
package-lifecycle expansion, covered by command semantics, and MCP's
protocol-mediated expansion, residual under command semantics. Both cases are
evaluated under the same run-level eligibility rule. An imported
execution enters the comparison only if it provides a captured approval record,
agent and version metadata, a non-prompt transcript or driver artifact,
observed-effect evidence, and artifact hashes. Each qualifying execution yields
a captured explicit-field represented set and an observed effect set. We count
an exact represented-to-observed mapping as recurrent across frontends only when
two such executions yield that same mapping. Prompt wording, local paths, and
model backend remain recorded run attributes rather than matching criteria;
command-aware and metadata-aware persistence are evaluated separately.

Applying this exact-mapping rule yields Finding~4.

\finding{4}{Two fixed-SHA exact mappings recur across three product
frontends.}{Across four qualifying runs, the Vite
P$\rightarrow$P,F and MCP M$\rightarrow$M,N mappings each recur under two
frontends. Vite resolves under command semantics, whereas MCP remains
command-residual; both resolve under the post hoc metadata-aware view.}

The four qualifying executions span tested configurations of Claude Code,
GitHub Copilot CLI, and Qwen Code.

For Vite, two approvals represent the pinned package-install process, and two
independent \path|node_modules/.modules.yaml| snapshots bind the additional
file effect. For MCP, two M-only records bind named, successful
documentation-tool calls and concrete responses; both executions additionally
exercise network authority.

These results provide the retrospective basis for asking
whether decision-time metadata can become a bound security object before the
approval decision.

\FloatBarrier

\section{Pre-Execution Recovery and Effect-Bound Records}
\label{sec:audit}

RQ3 turns retrospective recoverability into a prospective test: whether a
source-backed effect prediction can be computed, frozen, and bound before the
approval decision. We evaluate this transition from recovery to approval
through a prospective predictor, an effect-bound record, and a real-hook
integration experiment. The evaluation tests how much decision-time
metadata improves predictive effect coverage, how often the resulting policy
requests additional approval at the evaluated operating point, and whether the
enriched record integrates with an existing approval path.

\subsection{Prospective Protocol and Freeze Discipline}

The evaluation follows a development-and-freeze sequence. Parser v1.0 was
frozen before the controlled cohort and natural-project cohort A. We preserved
A's errors, used them to develop v1.1, and then froze v1.1, the evaluation
protocol, source hashes, rule hashes, and all B predictions before any
confirmatory execution. Each of the 18 primary B cases was run once under the
primary protocol; no primary outcome altered the parser or its predictions.
Appendices~\ref{app:supplementary-evidence} and~\ref{app:audit-diagnostics}
report the development scores and error profiles.

At the center of the frozen pipeline is a decision-time metadata parser. It
reads raw local
\path|package.json|, \path|pyproject.toml|,
\path|Cargo.toml| and \path|build.rs|, Dockerfile instructions, VS Code task
JSONC, MCP configuration, Git hooks, and referenced shell or PowerShell
scripts. It excludes benchmark-generated reports, execution traces,
observed-effect files, and harness instrumentation. For comparison, the
command-only baseline sees only the prompt-visible surface invocation and maps
generic command semantics to a conservative effect set.

For each target-executed workflow $i$, let
$E_i\subseteq\Omega_6$ be the independently observed effect set and
$\widehat E_i\subseteq\Omega_6$ the frozen prediction. We macro-average
case-level recall and precision equally across those workflows and report
exact-set accuracy as the fraction with $\widehat E_i=E_i$. Recall is one
when the observed set is empty; precision is one only when both sets are empty
and zero when exactly one is empty. An empty observed set does not enter the
coverage denominator and still receives these recall and precision scores.
A workflow that does not reach target execution stays in protocol accounting
and out of these outcome-dependent metrics.
Appendix~\ref{app:audit-metrics} gives the formulas.

\subsection{Prespecified Confirmatory Holdout B}

B prespecifies 18 workflows from 18 previously unseen fixed-SHA repositories:
one positive and one negative or low-risk control for each of npm, PEP~517,
Cargo, Docker, IDE tasks, MCP, Git hooks, nested shell, and unittest import.
Seventeen workflows reach target execution. Thirteen complete successfully,
and four terminate after producing effect evidence. One fails during setup
before the target workflow. The 17 target-executed cases form the
prediction-accuracy denominator; the setup failure remains in protocol and
terminal-flow accounting but supplies no realized effect set.

\begin{figure}[t]
\centering
\includegraphics[width=\columnwidth]{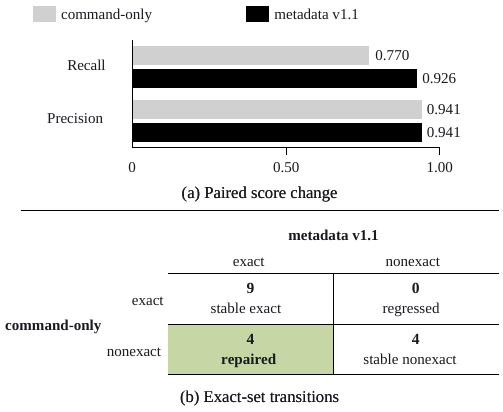}
\caption{Confirmatory-B paired comparison across the 17 target-executed cases.
The transition matrix tracks how command-only exact-set status changes under
frozen metadata v1.1.}
\label{fig:confirmatory-b-paired}
\end{figure}

On those 17 cases, positive recall and precision are 0.972 and 1.000; control
recall and precision are both 0.875. All three predeclared targets pass:
overall recall $\geq .80$, overall precision $\geq .80$, and executed-control
precision $\geq .80$. Both prediction sets, their freeze time, the parser/rule
hash, and the source hashes are bound in the pre-execution bundle.
Figure~\ref{fig:confirmatory-b-paired} reports the paired score change.

\finding{5}{Frozen metadata improves prospective recovery on the paired
holdout.}{Metadata raises macro recall from .770 to .926 at the same .941 macro
precision, while exact-set matches rise from 9/17 to 13/17 through four repairs
and zero exact-set regressions.}

\noindent\textbf{Independent effect-set reconstruction.}
A separately implemented verifier hash-checks 109 artifacts and reconstructs
all 17 target-executed effect sets independently of the predictor and original
collector labels, with zero effect-set or artifact-hash mismatches; it marks the
setup failure \path|NOT_EXECUTED|. Six prespecified cases rerun three times each
have mean pairwise effect-set Jaccard 1.000. Appendix
Table~\ref{tab:oracle-predicates} gives the scoring predicates;
Appendix~\ref{app:audit-diagnostics} provides the error localization and
sensitivity analyses.

\subsection{Effect-Bound Approval Record}

An effect-bound record commits the frozen prediction and its provenance before
authorization. Sandboxing constrains execution after the decision; the record
changes what the approval object represents before execution.

We instantiate the approval transaction model from Section~\ref{sec:model}.
Let $b_{\mathrm{cmd}}$ denote the prospective command-only object for the same
invocation. Because its represented and command-semantic sets are contained in
the frozen metadata prediction, the effect-bound object seals that complete set:
\begin{equation}
 \begin{aligned}
 p_{\mathrm{EB}}&=\mathrm{Seal}(b_{\mathrm{EB}},d),\\[-1pt]
 \mathrm{Rep}_6(p_{\mathrm{EB}})&=\widehat E_6
 =\mathrm{Pred}_6(a,m,s_0).
 \end{aligned}
 \label{eq:effect-bound-object}
\end{equation}
Before the decision, $b_{\mathrm{EB}}$ binds the invocation, frozen prediction,
metadata sources and hashes, parser/rule provenance, binding mode, and adapter
routing outcome; sealing adds the final permission outcome $d$. After a
permitted execution, transcript and realized-effect hashes are appended for
audit but cannot retroactively become approval-time disclosure. This
construction composes directly with
$\mathrm{CtxBind}$: in a closure-bound deployment, the mediator's binding layer
supplies the full transaction context, while the effect-bound record supplies
its effect-assertion component.

Having established what the record contains, we next test whether it can be
produced synchronously on the approval path. At the evaluated scale, the parser
satisfies that requirement. Across 12 controlled and 8 fixed-SHA
real-project metadata snapshots, every prediction is stable, with no parser
failure or timeout. Worst same-process P95 is
12.3\,ms; the all-case fresh-process maximum P95, including Python startup,
is 168.2\,ms. These measurements establish parser-level preflight cost; the
hook experiment below measures integrated adapter overhead.

\subsection{Approval-Path Integration and Policy Operating Point}

We next reuse the frozen B predictions in an approval-path policy to measure one
disclosure--selectivity operating point. The policy returns routing outcome
$h=\textsc{ASK}$, requesting an additional approval decision, when metadata
identifies induced mechanisms, added effects, or warnings. \textsc{ASK}
delegates to the existing permission flow; it is not the final outcome $D(p)$.
If metadata produces no signal, the request passes through to that flow.
Invalid output fails closed, and the policy never auto-allows.
We operationalize demand for additional approvals as the fraction of
prespecified cases routed to \textsc{ASK}, directly measuring how often this
policy invokes another decision at the product boundary.

To quantify disclosure independently of the policy consumer, let
$E_i=\mathrm{Eff}_6(\tau_i)$ and let $R_i$ be the effect set disclosed by the
evaluated record. For an effect-bound record, instantiating
\eqref{eq:effect-bound-object} per workflow gives
$R_i=\widehat E_{6,i}=\mathrm{Rep}_6(p_{\mathrm{EB},i})$ once sealed. Equation~\eqref{eq:coverage-residual} in
Appendix~\ref{app:audit-metrics} averages coverage over the 16 target-executed
cases with nonempty $E_i$ and counts residual instances over all 17. The
remaining control has $E_i=\emptyset$ and $R_i\ne\emptyset$, so its coverage
ratio is undefined. It is not an exact-set match and contributes zero residuals.

Table~\ref{tab:approval-path-tradeoff} reports the resulting operating point,
separating overall disclosure from the positive and control strata and placing
the corresponding policy-routing demand alongside each stratum.

\begin{table}[!t]
\caption{Confirmatory-B disclosure and routing by stratum.}
\label{tab:approval-path-tradeoff}
\centering
\footnotesize
\setlength{\tabcolsep}{2.4pt}
\begin{tabular}{@{}lccccc@{}}
\toprule
Scope & \shortstack{Coverage\\Cmd.$\to$EB} &
\shortstack{Residual\\Cmd.$\to$EB} &
\shortstack{Exact sets\\Cmd.$\to$EB} &
\shortstack{ASK\\protocol} &
\shortstack{No-added-\\effect ASK} \\
\midrule
Overall  & .755$\to$.922 & 10$\to$3 & 9/17$\to$13/17 & 17/18 & 9/17 \\
Positive & .676$\to$.972 & 8$\to$1  & 4/9$\to$8/9     & 9/9   & 4/9 \\
Controls & .857$\to$.857 & 2$\to$2  & 5/8$\to$5/8     & 8/9   & 5/8 \\
\bottomrule
\end{tabular}
\par\vspace{2pt}
\begin{minipage}{\columnwidth}
\raggedright
\emph{Note.}
Cmd. and EB denote command-only and effect-bound records, respectively.
Denominators are overall/positive/control: 16/9/7 for coverage over nonempty
observed effect sets; 17/9/8 for residual and exact-set results over
target-executed cases; 18/9/9 for ASK routing over all protocol cases; and
17/9/8 for no-added-effect ASK over reached cases. A no-added-effect ASK is a
reached case routed to ASK whose realized trace exercises no effect class
beyond those disclosed by the command-only record.
\end{minipage}
\end{table}

The seven newly disclosed instances are in the positive stratum, and control
disclosure is unchanged. Routing demand is separate from that record-level
gain: the evaluated policy sends 9/17 reached cases to \textsc{ASK} with no
added effect class, with one warning and no parser failures.

\finding{6}{Effect-bound records disclose seven additional effect
instances.}{Reusing the frozen predictions raises case-macro coverage from
.755 to .922 and reduces residual effect instances from 10 to 3. The evaluated
conservative policy makes the disclosure--selectivity tradeoff explicit.}

\noindent\textbf{Exploratory offline selectivity replay.}
Only the frozen metadata-signal rule was predeclared. An exploratory offline
replay of three additional rules on the same frozen predictions and realized
traces leaves disclosure unchanged. Restricting \textsc{ASK} to added effects
or warnings lowers protocol demand from 17/18 to 7/18 and leaves 2/7
command-residual cases without escalation
(Appendix Table~\ref{tab:ask-selectivity}).

The B operating point measures cohort-level policy behavior; we next exercise
insertion of the same frozen record into an existing product approval path. In
the predeclared npm case, 100 in-process preflights
have P95 1.4\,ms; 30 fresh hook processes have median 81.4\,ms and P95
84.8\,ms, including startup, hook input, parsing, and serialization. An
isolated Claude Code 2.1.205 \path|PreToolUse| integration run then issues the
exact \texttt{npm run build} request. The hook emits $h=\textsc{ASK}$, the
noninteractive permission path returns final outcome $d=\textsc{DENY}$, no
automatic allow occurs, and the workspace remains hash-identical. This run
shows that the adapter can place the frozen record on the product permission
path. In this run the adapter binds the predicted effects into an ASK routing outcome.

\subsection{Controlled Decision Pair}

The holdout above compares predicted labels. It does not hold one entry
command and one effect policy fixed while an expansion source changes. We
therefore ran one pair outside Confirmatory~B, with parser v1.1 left
unmodified. Both workspaces use the same \path|package.json| and the surface
command \texttt{npm run build}. The pre-approval difference is \path|build.js|:
the local script writes a file, and the network script writes that file and
then calls \texttt{fetch} on a loopback URL. The researcher policy allows
process and file effects and forbids network. Command semantics for this
command are \(\{\texttt{process}\}\) on both arms. Both arms are executed for
measurement. Table~\ref{tab:decision-pair} reports that pair.

\setlength{\textfloatsep}{8pt}
\begin{table}[!t]
\caption{One controlled pair under a fixed command and a fixed researcher
effect policy. Modeled and study-policy rows are computed before measurement;
measured rows come from independent execution of both arms.}
\label{tab:decision-pair}
\centering
\footnotesize
\setlength{\tabcolsep}{3pt}
\begin{tabular}{@{}lcc@{}}
\toprule
View & Local script & Network script \\
\midrule
Modeled command-approval baseline & ALLOW & ALLOW \\
Command semantics & ALLOW & ALLOW \\
Mechanism-triggered ASK & ASK & ASK \\
Metadata policy (study) & ALLOW & DENY \\
\midrule
Measured effects & process, file & process, file, network \\
Local task completed & yes & yes \\
Loopback request observed & no & yes \\
\bottomrule
\end{tabular}
\end{table}

A study
metadata gate on the same pair prevents launch after a metadata denial. Five
\path|build.js| rewrites of the network arm then yield 3 correct blocks and 2
erroneous allows with loopback contact. The command-approval baseline is
modeled, the gate is not a product-native dialog, and the stress result is
limited to this five-variant set.

\FloatBarrier
\setlength{\textfloatsep}{1.55\baselineskip plus 0.2\baselineskip minus 0.4\baselineskip}

\section{Discussion}

\noindent\textbf{Guarantee boundary.}
The proved statement is conditional. When two validly bound executions present
the same complete policy-visible input and a pre-specified effect policy
requires different decisions, no record-only rule guarantees both. The
controlled pair and five-variant stress set are researcher constructions: the
command-approval baseline is modeled, and the gate is not a product-native
dialog. Confirmatory B prespecifies no environment-positive target, and its
outcome collector does not measure environment effects.
\(\Omega_6\) remains a coverage vocabulary for class labels. It does not
express path-sensitive file policies, destination identity, or a product's
deployed command contract.

\noindent\textbf{Security implication.}
A truthfully identified invocation remains insufficient when the decision-time
approval object omits effects activated by its workflow, so the repair is to
bind that effect boundary before authorization.

\noindent\textbf{Generalization and design principle.}
An approval object should bind a source-backed effect boundary before
authorization. This relation applies where systems expose an operational effect
vocabulary and inspectable expansion sources, together with evidence of
activation and realized effects.

\section{Related Work}

Prior work addresses neighboring security objects across agent execution,
authorization, and workflow expansion. We organize these lines by the property
they constrain.

\textbf{Agent execution and tool security.}
XTHP studies malicious companion tools that redirect selection and control flow
in tool pools~\cite{xthp}. IsolateGPT isolates application execution, while ACE
constrains planning and execution under information-flow and capability
policies~\cite{isolategpt,ace}. PFI isolates a trusted agent from an untrusted agent and keeps untrusted data inside the restricted agent~\cite{pfi}. ToolGuardian derives facts from descriptions,
traces, mock execution, and source analysis for tool admission and runtime
authorization~\cite{toolguardian}. Adjacent work studies prompt injection,
task-context projection for agent-to-agent messages, tool-invocation behavior
of coding agents, tool emulation, excessive agency, coding-assistant behavior,
permission gates, and agent-human interaction
~\cite{indirectpromptinjection,imprompter,agentfall,agentfirewalls,redteamtoolinvocation,toolemu,agentdojo,owaspexcessiveagency,
perryinsecurecode,permissiongate,agenthumaninteraction}. We instead hold the
selected entry invocation
fixed and faithfully identified, then measure whether its decision-time
approval object covers the transitive operational effects activated by the
selected implementation.

\textbf{Authorization scope, action identity, and provenance.}
Classic authorization establishes complete mediation, exposes confused-deputy
risks, and constrains delegation
~\cite{saltzerschroeder,hardyconfuseddeputy,macaroons,
overlayinggovernance}. Prior work on cross-layer systems shows that individually
meaningful authorization policies can fail to compose across software
layers~\cite{georgiev2014breaking}. Approval laundering exposes a record-level
form of this broader compositional hazard: even when entry-invocation identity
is preserved, the durable decision-time record may not cover the selected
implementation's execution closure. Permission studies examine whether
authorization boundaries are intelligible and how agents request and enforce
user permission~\cite{feltpermissions,androidremystified,agentpermissionssurvey,smartauth,conseca}. Janus compares user-involved designs for agent permission decisions~\cite{janus}. Recent agent
work protects adjacent properties. SkillScope applies fine-grained least
privilege to agent skills~\cite{skillscope}. AuthGraph compares an
authorization graph taken from user intent with later execution
provenance~\cite{authgraph}. PAuth authorizes an operation together with
its operands~\cite{pauth}. SEAgent enforces a mandatory policy that can
prompt the user~\cite{seagent}. PACT checks each tool argument's origin
against its role trust contract~\cite{pact}. ConLeash binds an MCP tool consent to a boundary and escalates a call that crosses it~\cite{conleash}. Vigil checks an execution trace against a skill policy for access and side effects~\cite{vigil}. ClawLess intercepts system calls to enforce an agent's permissions and trust scopes~\cite{clawless}. Progent checks each tool call against rules over its name and arguments, and requires approval before those rules expand~\cite{progent}. AuthBench infers task-level file policies
and evaluates execution-closure proxies~\cite{authbench}, while Consent
Integrity binds approval to the exact boundary action rendered by a trusted
mediator~\cite{consentintegrity}. The distinction is direct: AuthBench asks
which authority a task requires; Consent Integrity asks whether the approved
action is the action that executes; we ask whether the decision-time approval
object for a faithfully identified invocation covers the transitive operational
effects activated by its selected implementation. Studies of authority,
identity, provenance, or delegated flow structure those properties
~\cite{overeager,schmotzskills,authorizationexecutiongap,pcaactions,cava,chaincaps,approvalframeddelegation,saga,ghostapproval}.

\textbf{Transitive expansion sources and approval envelopes.}
Studies of package-install and dependency attacks constrain the expansion source~\cite{smallworld,maloss,latch}.
Studies of agent skills, MCP tool calls, and approval-request profiles constrain the envelope presented at the decision~\cite{claudecodeskills,githubagentskills,mcp2026spec,authzenaarp,authzencoazmcp,openaiagentshitl}.

\section{Conclusion}

Approval laundering is a coverage gap in a durable approval record: the record
identifies an entry invocation and omits an effect of the workflow that
invocation starts. The gap is an authorization failure when a pre-specified
effect policy requires different decisions and the complete policy-visible
inputs are identical. Under that condition, no record-only policy can
guarantee both decisions. Across the measured mechanisms and fixed-SHA
projects, the explicit, command, and metadata views locate where labels are
missing. On the prespecified holdout, binding frozen predictions cuts residual
effect instances from 10 to 3. Across the 40 explicit coverage gaps, 8 have
driver-observed permits and 0 are structural witnesses once full context
binding is required. We privately disclosed the product-specific
approval-record findings for Claude Code, GitHub Copilot CLI, and Qwen Code
to Anthropic, GitHub, and Alibaba, with version-specific reproduction
instructions and supporting evidence.

\section*{Ethics Considerations}

\textbf{Study conduct.}
This defensive study executes controlled fixtures and public fixed-SHA
repositories in isolated workspaces. It involves no human subjects or personal
data and accesses no private repository or unrelated production target.
Authentication uses researcher-controlled accounts; credential checks use
synthetic canaries or non-reversible evidence. Network activity is limited to
ordinary product access and task-bounded calls to a documented public MCP
endpoint. The experiments neither submit upstream changes nor generate traffic
intended to affect maintainers or users.

\textbf{Responsible disclosure.}
We privately disclosed the product-specific approval-record findings for Claude
Code, GitHub Copilot CLI, and Qwen Code to Anthropic, GitHub, and Alibaba,
respectively. Each disclosure included version-specific reproduction
instructions and supporting evidence to facilitate independent validation.

\textbf{Authorized use and dual use.}
The controlled demonstrations have dual-use potential and are intended only for
authorized reproduction, defensive evaluation, and improvement of agent approval
mechanisms. The paper and artifact do not grant authorization to probe, exploit,
or disrupt third-party systems; reproduction should be limited to systems the
evaluator owns, controls, or has explicit permission to test. They do not
authorize any form of unlawful real-world attack, or any application of these
demonstrations outside that limit.

\textbf{Deployment responsibility.}
Effect-bound approval can create harm if incomplete predictions produce false
assurance, over-broad \textsc{ASK} routing produces approval fatigue, or durable
records expose repository and infrastructure metadata. Deployers should retain
accountable policy ownership, minimize secret-bearing content before sealing,
and govern records as sensitive security telemetry through access and retention
controls.

\section*{Acknowledgments}
Generative AI was used for editorial purposes in this manuscript, and all
outputs were inspected by the authors to ensure accuracy and originality.

\bibliographystyle{IEEEtran}
\bibliography{references}

\appendices
\raggedbottom
\section{Supplementary Experimental Evidence}
\label{app:supplementary-evidence}

This appendix provides the detailed evidence tables, denominator accounting,
and claim-level analyses referenced from the main paper.

\subsection{Claim Evidence and Denominator Accounting}

Across the 111 canonical records selected under the latest-success rule, the
experiment summary contains 81 trace-reported effect-label assignments. An
alternative first-success rule changes only three agent--task keys and yields
39/111. The
construct-validity audit excludes inherited-read \texttt{E} labels from 68 records,
fixture-authored protocol-shaped \texttt{M} labels from 11 records, and strict
\texttt{M}/\texttt{P} labels from the \path|MCP_NOT_VISIBLE| attempt. These
operations produce 41 trace-preserving label reclassifications without
modifying a ledger row or run snapshot.

The first-success sensitivities of the command and metadata views are 16/111
and 12/111.

\begin{table}[!b]
\caption{Mechanism-level residual counts across three RQ1 representation
views.}
\label{tab:rq1-mechanism-effects}
\centering
\begin{tabular}{@{}lrrrr@{}}
\toprule
Mechanism & $n$ & Explicit & +Command & +Metadata \\
\midrule
Cargo build script & 15 & 5 & 0 & 0 \\
Docker build & 2 & 2 & 0 & 0 \\
Git hook & 11 & 3 & 0 & 0 \\
IDE task & 13 & 4 & 3 & 3 \\
MCP server & 14 & 6 & 6 & 4 \\
npm lifecycle & 17 & 7 & 3 & 2 \\
Python PEP~517 & 15 & 6 & 0 & 0 \\
Nested shell & 13 & 4 & 3 & 2 \\
Test-runner import & 11 & 3 & 2 & 2 \\
\midrule
\textbf{Total} & \textbf{111} & \textbf{40} & \textbf{17} & \textbf{13} \\
\bottomrule
\end{tabular}
\par\vspace{2pt}
\begin{minipage}{\columnwidth}
\raggedright
\emph{Note.}
All three views use the same 111 canonical records. Explicit uses captured
approval fields; +Command adds generic command semantics; +Metadata applies the
frozen v1.1 parser post hoc to raw metadata available at decision time.
\end{minipage}
\end{table}

\begin{table}[H]
\caption{RQ1 denominator and ledger accounting.}
\label{tab:evidence-state}
\centering
\footnotesize
\begin{tabular}{@{}>{\raggedright\arraybackslash}p{0.57\columnwidth}r
>{\raggedright\arraybackslash}p{0.22\columnwidth}@{}}
\toprule
Unit & Count & Selection \\
\midrule
Execution attempts / run snapshots & 149 & Ledger total \\
\path|SUCCESS| terminal snapshots & 139 & 10 non-success \\
\textbf{Deduplicated successful agent--task records} & \textbf{111} &
\textbf{Latest success/key} \\
\bottomrule
\end{tabular}
\end{table}

\begin{table}[!t]
\caption{Structural-witness eligibility funnel on the 111 canonical RQ1
records. $G$ is an explicit coverage gap; $\mathrm{Permit}$ requires a
same-run allowing decision for the exact invocation; $\mathrm{Bind}$ requires
both context binding and trace attribution. Absent historical evidence is
\texttt{insufficient}.}
\label{tab:witness-funnel}
\centering
\footnotesize
\begin{tabular}{@{}lrrrr@{}}
\toprule
Source stratum & $N$ & $G$ & $G\land\mathrm{Permit}$ & $W$ \\
\midrule
Harness/parser fixtures & 60 & 30 & 0 & 0 \\
Cross-agent fixtures & 40 & 0 & 0 & 0 \\
Fixed-SHA project executions & 11 & 10 & 8 & 0 \\
\midrule
Total & 111 & 40 & 8 & 0 \\
\bottomrule
\end{tabular}
\end{table}

The separate 111-task annotation-backed reference supplies an independent
comparison for this representation ladder. Prompt-only recall is 0.616;
annotation replay gives recall 1.000, precision 0.976, eight false-positive
rows, and 0.072 additional predictions per row, with none among 22 controls.
Expected-effect annotations make this the annotation-backed reference, while
headline estimates use the frozen non-annotation views. Its provenance records
the parser and rule versions, prediction time, source and outcome hashes,
leakage status, collector, terminal state, and artifact-predicate recomputation.

\begin{table*}[!t]
\caption{Post-approval evidence for the 11 fixed-SHA executions in
Table~\ref{tab:real-projects}; the Vite and MCP remote-call rows each aggregate
two runs.}
\label{tab:fixed-sha-evidence}
\centering
\footnotesize
\setlength{\tabcolsep}{4pt}
\rowcolors{2}{black!4}{white}
\begin{tabular}{@{}>{\raggedright\arraybackslash}p{0.285\textwidth}c
>{\raggedright\arraybackslash}p{0.625\textwidth}@{}}
\toprule
Case, short SHA, and execution view & Runs & Post-approval evidence \\
\midrule
\path|webpack/webpack| \texttt{@efeb5ce2} & 1 &
transcript, file delta, and Husky-created \path|.husky/_| shims \\
\path|vitejs/vite| \texttt{@64dfee12} & 2 &
two agent transcripts and two \path|node_modules/.modules.yaml| snapshots \\
\path|pypa/pip| \texttt{@72e6c594} & 1 &
file delta and generated wheel \\
\path|pallets/flask| \texttt{@36e4a824} & 1 &
file delta and generated example wheel \\
\path|BurntSushi/ripgrep| \texttt{@dfe4a81d} & 1 &
transcript and \path|target/debug/rg.exe| \\
\path|microsoft/vscode| \texttt{@8d5908d6} & 1 &
iidfile, image inspection, and image history \\
\path|docker/awesome-compose| \texttt{@30f4b7f6} & 1 &
iidfile and Docker image inspection \\
modelcontextprotocol/\allowbreak servers \texttt{@7b1170d1} (configuration) & 1 &
\path|MCP_NOT_VISIBLE|; no activated capability or post-approval child process \\
modelcontextprotocol/\allowbreak servers \texttt{@7b1170d1} (remote calls) & 2 &
named documentation-tool calls and concrete responses \\
\bottomrule
\end{tabular}
\end{table*}

RQ2 uses the fixed-SHA recurrence denominator rather than the
construct-control cells.

Separate artifact-only npm, Cargo, and MCP checks provide three passing
external trace validations without fixture event logs.
Table~\ref{tab:agent-configurations} separates evaluated cells from refusals
and configurations that were not scheduled.

The prospective sequence moves from the controlled cohort to Natural A and
then to confirmatory B. The controlled cohort contains 12 workflows, including
six controls, and yields macro recall 0.903 and macro precision 0.917. Natural A
contains 12 workflows from nine repositories, including six controls, and
yields recall 0.694 and precision 0.732.
The v1.0 development cohorts remain separate from the frozen v1.1
confirmation.

Frozen hashes bind the predictions and inputs for every cohort.
Appendix~\ref{app:audit-diagnostics} supplies the class-level error profiles,
terminal-state sensitivities, and repeated-run stability analysis.

\begin{table}[!t]
\caption{Shared-fixture cross-frontend construct control.}
\label{tab:agent-configurations}
\centering
\footnotesize
\setlength{\tabcolsep}{2.5pt}
\begin{tabular}{@{}lcccc@{}}
\toprule
Configuration & Version & \shortstack{Evaluable/\\Scheduled} & Refused & Unscheduled \\
\midrule
Claude Code (primary) & 2.1.193 & 8/8 & 0 & 0 \\
Claude Code (Gemma, supp.) & 2.1.193 & 2/2 & 0 & 6 \\
Codex CLI & 0.142.0 & 8/8 & 0 & 0 \\
GitHub Copilot CLI & 1.0.65 & 8/8 & 0 & 0 \\
OpenCode & 1.1.25 & 6/8 & 2 & 0 \\
Qwen Code & 0.19.2 & 8/8 & 0 & 0 \\
\midrule
Total & --- & 40/42 & 2 & 6 \\
\bottomrule
\end{tabular}
\par\vspace{2pt}
\begin{minipage}{\columnwidth}
\raggedright
\emph{Note.}
All 40 evidence-complete cells yield zero strict gaps.
\end{minipage}
\end{table}

\begin{table}[!t]
\caption{Confirmatory-B raw-artifact predicates (Oracle v1.0.1).}
\label{tab:oracle-predicates}
\centering
\footnotesize
\begin{tabular}{@{}p{0.19\columnwidth}p{0.73\columnwidth}@{}}
\toprule
Effect & Raw-artifact predicate \\
\midrule
Process & At least one process-observation step, or a successful MCP stdio
exchange. \\
File & A workspace or sandbox delta contains a created, modified, or deleted
path; Git HEAD changes; or a Docker image is successfully inspected. \\
Network & At least one captured network connection, or a successful MCP
Streamable HTTP exchange. \\
Container & Successful Docker inspection with a nonempty image identifier. \\
MCP & \path|mcp_protocol.json| records at least one successful stdio or
Streamable HTTP server exchange. \\
Environment & Not measured by the confirmatory-B outcome collector; B
prespecifies no environment-positive target. \\
\bottomrule
\end{tabular}
\par\vspace{2pt}
\begin{minipage}{\columnwidth}
\raggedright
\emph{Note.}
All 109 raw artifacts pass hash verification. A separately implemented verifier
exactly reconstructs the effect sets of all 17 target-executed cases, marks the
remaining setup-failure case as \texttt{NOT\_EXECUTED}, and reports no
effect-set or artifact-hash mismatches.
\end{minipage}
\end{table}

The parser microbenchmark contains 12 controlled and eight selected-project
inputs. After five warmups, each is evaluated in 100 same-process repetitions
and 30 fresh processes. Every prediction digest is stable, with no failure or
timeout; the worst same-process P95 is 12.298\,ms, and the selected-project and
all-case fresh-process maximum P95 values are 164.752 and 168.241\,ms.

Exact-set accuracy and control precision vary most across the reported
confirmatory-B profiles.
Table~\ref{tab:prospective-effect-errors} localizes the cohort errors.

\subsection{Proof of Lemma 1}
\label{app:lemma-proof}

By~\eqref{eq:record-only-indistinguishability}, both instances induce the same
output distribution $\mu$. Since $q_1\ne q_2$, the singleton events $\{q_1\}$
and $\{q_2\}$ are disjoint, so $\mu(\{q_1\})+\mu(\{q_2\})\le 1$.

\begingroup
\raggedbottom

\section{Secondary Audit Analyses}
\label{app:audit-diagnostics}

These analyses add class-level error localization, terminal-state sensitivity,
repeated-run stability, and runtime detail while preserving the primary
estimates in Section~\ref{sec:audit}.

\subsection{Metric Definitions and Denominators}
\label{app:audit-metrics}

For each target-executed workflow $i$, let $E_i\subseteq\Omega_6$ and
$\widehat E_i\subseteq\Omega_6$ denote the independently observed and frozen
predicted effect sets.
\noindent Section~\ref{sec:audit} uses the following case-level scores.
\begin{equation}
\begin{aligned}
\rho_i&=
\begin{cases}
1, & E_i=\emptyset,\\
|\widehat E_i\cap E_i|/|E_i|, & \text{otherwise},
\end{cases}\\
\pi_i&=
\begin{cases}
1, & \widehat E_i=E_i=\emptyset,\\
|\widehat E_i\cap E_i|/|\widehat E_i|, & |\widehat E_i|>0,\\
0, & \text{otherwise}.
\end{cases}
\end{aligned}
\label{eq:case-recall-precision}
\end{equation}
Macro recall and precision average the case-level scores
in~\eqref{eq:case-recall-precision} equally over the target-executed workflows
in each cohort; the primary Confirmatory-B denominator is 17. Its one additional
workflow that does not reach target execution remains in protocol accounting
but not in these outcome-dependent metrics.

For a disclosed effect set $R_i$, 16 of the 17 target-executed cases have a
nonempty observed set. Case-macro coverage and residual effect instances are
\begin{equation}
 \begin{aligned}
 \mathrm{Cov}(R)&=\textstyle\frac{1}{16}\sum_{i:\,|E_i|>0}
 \textstyle\frac{|R_i\cap E_i|}{|E_i|},\\
 \mathrm{Residual}(R)&=\textstyle\sum_{i=1}^{17}|E_i\setminus R_i|.
 \end{aligned}
 \label{eq:coverage-residual}
\end{equation}
Table~\ref{tab:ask-selectivity} replays routing rules on the same frozen
predictions and realized traces.

\begin{table}[!ht]
\caption{Offline ASK selectivity on Confirmatory~B. Disclosure
(.755$\rightarrow$.922 coverage; 10$\rightarrow$3 residuals) is unchanged
across rows. ASK/18 is protocol routing demand; no-added-effect ASK uses the
17 reached cases; cmd-residual columns use the 7 reached cases whose
command-only residual is nonempty.}
\label{tab:ask-selectivity}
\centering
\footnotesize
\setlength{\tabcolsep}{2.2pt}
\begin{tabular}{@{}lcccc@{}}
\toprule
Routing rule & ASK/18 & Pass/18 &
\shortstack{No-added\\ASK} &
\shortstack{ASK covers\\cmd residual} \\
\midrule
Frozen metadata signal$^\dagger$ & 17 & 1 & 9/17 & 7/7 \\
Added-effect/warning$^\ast$ & 7 & 11 & 2/17 & 5/7 \\
Network/added/warning$^\ast$ & 9 & 9 & 4/17 & 5/7 \\
Command pass-through$^\ast$ & 0 & 18 & 0/17 & 0/7 \\
\bottomrule
\end{tabular}
\par\vspace{2pt}
\begin{minipage}{\columnwidth}
\raggedright
\emph{Note.}
$^\dagger$Predeclared frozen policy.
$^\ast$Exploratory offline replay on frozen predictions.
No-added-effect ASK is a post-execution proxy.
Pass-through with a nonempty command residual is an under-selective proxy
(2, 2, and 7 reached cases for the three starred rows).
\end{minipage}
\end{table}

\subsection{Prospective Cohort Error Profiles}

Controlled v1.0 pairs one positive with one control in each of six classes;
class recall/precision are Cargo 1.000/0.833, IDE 0.833/1.000, MCP 0.583/1.000,
npm 1.000/1.000, PEP~517 1.000/0.667, and shell 1.000/1.000.

\looseness=-1
Natural A has ten successful executions and two evidence-complete command
failures. Exact-set accuracy is 0.167; positive and control recall/precision are
0.778/0.819 and 0.611/0.644. The successful-only view gives
0.733/0.678/0.200. Five missed file effects and five extra environment
predictions dominate; the post hoc surface/metadata partition localizes them to
the corresponding prediction layer.

Across the 17 target-executed Confirmatory B cases, three cases miss a file
effect, while one process effect and one MCP effect are extra.

\begin{table}[!ht]
\caption{Effect-class errors across prospective cohorts. Entries are cases
with missed (M) or extra (E) effects; gray zeros denote no cases.}
\label{tab:prospective-effect-errors}
\centering
\footnotesize
\setlength{\tabcolsep}{2.7pt}
\newcommand{\zv}{\textcolor{black!45}{0}}
\begin{tabular}{@{}lrrrrrr@{}}
\toprule
& \multicolumn{2}{c}{\shortstack{Controlled\\[-1pt]{\footnotesize $n=12$}}}
& \multicolumn{2}{c}{\shortstack{Natural A\\[-1pt]{\footnotesize $n=12$}}}
& \multicolumn{2}{c}{\shortstack{Confirmatory B\\[-1pt]{\footnotesize $n=17$}}} \\
\cmidrule(lr){2-3}\cmidrule(lr){4-5}\cmidrule(l){6-7}
Effect & M & E & M & E & M & E \\
\midrule
Process     & \zv & \zv & \textbf{1} & \zv & \zv & \textbf{1} \\
File        & \textbf{3} & \zv & \textbf{5} & \textbf{1} & \textbf{3} & \zv \\
Environment & \zv & \textbf{3} & \zv & \textbf{5} & \zv & \zv \\
Network     & \zv & \zv & \textbf{2} & \textbf{2} & \zv & \zv \\
Container   & \zv & \zv & \zv & \textbf{1} & \zv & \zv \\
MCP         & \zv & \zv & \zv & \zv & \zv & \textbf{1} \\
\bottomrule
\end{tabular}
\end{table}

\subsection{Confirmatory-B Sensitivities}

Confirmatory B contains 18 prespecified cases: 13 successes, four
evidence-complete command failures, and one setup failure. The primary
target-executed view reports 0.875 control recall/precision. Treating
\path|NOT_EXECUTED| as empty yields recall/precision/exact-set accuracy
0.931/0.889/0.722 and 0.778 control precision; restricting to successes yields
0.942/0.923/0.769 and 0.833.

Six predeclared three-run cases have mean pairwise Jaccard 1.000. Excluding
Node/V8 cache deltas yields recall/exact-set accuracy 1.000/0.889. For the one
target-executed empty-observed case, scoring recall as one, excluded, or zero
yields 0.926/0.922/0.868; all exceed 0.80. Precision (0.941) and exact sets
(13/17) are unchanged.
\endgroup

\end{document}